\documentclass{stavanger}

\usepackage[utf8]{inputenc}
\usepackage{amssymb,amsmath,amsfonts,amsthm}
\usepackage{hyperref}
\usepackage{natbib}
\usepackage{graphicx}
\usepackage{tikz}
\usepackage{placeins}

\startlocaldefs

\endlocaldefs

\begin{document}

%%%%%%%%%%%%%%%%%%%%%%%%%%%%%%%%%%%%%%%%%
%%                                     									       %%
%%  Start of the title page information section						       %%
%%                                     									       %%
%%%%%%%%%%%%%%%%%%%%%%%%%%%%%%%%%%%%%%%%%

\begin{frontmatter}

\begin{fmbox}

%%%%%%%%%%%%%%%%%%%%%%%%%%%%%%%%%%%%%%%%%
%%                                     									       %%
%%  Set the header on the first page, default is "Research Article"		       %%
%%                                     									       %%
%%  Define which division of the Institute of Mathematics and Physics	       %%
%%  this work is associated with								       %%
%%                                     									       %%
%%        PM:	Pure Mathematics                         					       %%
%%        ST:	Mathematical Statistics                     					       %%
%%        MP:	Mathematical Physics                        					       %%
%%        MS:	Material Physics                         					       %%
%%        FP:	Theoretical Physics							       	       %%
%%                                     									       %%
%%%%%%%%%%%%%%%%%%%%%%%%%%%%%%%%%%%%%%%%%

\dochead{Research Article }{FP}

%%%%%%%%%%%%%%%%%%%%%%%%%%%%%%%%%%%%%%%%%
%%                                     									       %%
%%  Title of the manuscript									       %%
%%                                     									       %%
%%%%%%%%%%%%%%%%%%%%%%%%%%%%%%%%%%%%%%%%%

\title{Reducing the Sign Problem with Line Integrals}

%%%%%%%%%%%%%%%%%%%%%%%%%%%%%%%%%%%%%%%%%
%%                                     									       %%
%%  Author information										       %%
%%                                     									       %%
%%%%%%%%%%%%%%%%%%%%%%%%%%%%%%%%%%%%%%%%%

\author[
   %addressref={aff1},                   	  % id's of addresses
   %corref={aff1},                     		  % id of corresponding address, if any
   %noteref={n1},                        		  % id's of article notes, if any
   %email={rasmus.n.larsen@uis.no}   		  % email address
]{\inits{FL}\fnm{Rasmus N.} \snm{Larsen}}

%%%%%%%%%%%%%%%%%%%%%%%%%%%%%%%%%%%%%%%%%
%%                                     									       %%
%%  Affiliations of the authors									       %%
%%                                     									       %%
%%%%%%%%%%%%%%%%%%%%%%%%%%%%%%%%%%%%%%%%%

\address[id=aff1]{%                          			 % unique id
  \orgname{Faculty of Science and Technology}, 	 % faculty
  \street{University of Stavanger},                     		 % university
  \postcode{4021}                               			 % post or zip code
  \city{Stavanger},                              				 % city
  \cny{Norway}                                   				 % country
}

%%%%%%%%%%%%%%%%%%%%%%%%%%%%%%%%%%%%%%%%%
%%                                     									       %%
%%  Additional notes on the authors								       %%
%%                                     									       %%
%%%%%%%%%%%%%%%%%%%%%%%%%%%%%%%%%%%%%%%%%

%\begin{artnotes}
%\note[id=n1]{Only contributor} % note, connected to author
%\end{artnotes}

\end{fmbox}% comment this for two column layout

%%%%%%%%%%%%%%%%%%%%%%%%%%%%%%%%%%%%%%%%%
%%                                     									       %%
%%  Abstract of the manuscript									       %%
%%                                     									       %%
%%%%%%%%%%%%%%%%%%%%%%%%%%%%%%%%%%%%%%%%%

\begin{abstractbox}

\begin{abstract}
We present a novel strategy to strongly reduce the severity of the sign problem, using line integrals along paths of changing imaginary action. Highly oscillating regions along these paths cancel out, decreasing their contributions. As a result, sampling with standard Monte-Carlo techniques becomes possible in cases that otherwise require methods taking advantage of complex analysis, such as Lefschetz-thimbles or Complex Langevin. We lay out how to write down an ordinary differential equation for the line integrals. As an example of its usage, we apply the results to a 1d quantum mechanical anharmonic oscillator with a $x^4$ potential in real time, finite temperature.  
\end{abstract}

%%%%%%%%%%%%%%%%%%%%%%%%%%%%%%%%%%%%%%%%%
%%                                     									       %%
%%  Keywords for the article, each one in its separate \kwd{}			       %%
%%                                     									       %%
%%%%%%%%%%%%%%%%%%%%%%%%%%%%%%%%%%%%%%%%%

%\begin{keyword}
%\kwd{keyword1}
%\kwd{keyword2}
%\kwd{keyword3}
%\end{keyword}

\end{abstractbox}

%\end{fmbox}% uncomment this for twcolumn layout

\end{frontmatter}

%%%%%%%%%%%%%%%%%%%%%%%%%%%%%%%%%%%%%%%%%
%%                                     									       %%
%%  Main text starts here										       %%
%%                                     									       %%
%%%%%%%%%%%%%%%%%%%%%%%%%%%%%%%%%%%%%%%%%

\section{Reducing the sign problem with line integrals}\label{sec:line}
The sign problem for integrals appears in models where the integrand is not positive definite, such as real time quantum mechanics or finite density quantum field theories. The problem lies in that large cancellations occur, such that even though the integrand contains large contributions, for example of order $10^3$, the sum conspires to give a result of order 1, or smaller. This might not seem too bad, but standard approaches like Monte-Carlo sampling errors behave as $1/\sqrt{N_s}$ where $N_s$ is the number of independent samples. If one samples values of order $10^3$, one would need 1 million independent samples to achieve the precision of the order of the result (and assuming that the errors should be around $1\%$, an extra factor of 10000). This problem is of course not new, and people have tried several approaches that often take advantage of analytic continuation into the complex plane. Examples of these are the  Lefschetz-thimbles method \cite{Cristoforetti:2012su,Alexandru:2016gsd} which uses a steepest descent approach to sample around thimbles of constant imaginary action, or Complex Langevin \cite{Seiler:2017wvd,Berges:2006xc,Alvestad:2021hsi} which derives an evolution equation for the expectation values that should be equivalent to the original distribution's expectation values.

In this paper, we propose a qualitatively different approach that does not require analytic continuation (though future work might include attempts to combine these). We wish to decrease the severity of the sign problem as much as possible. Our starting point is the observation that for an integral I with an observable O
\begin{eqnarray}
I(O) &=& \int d^N x O(x)\exp(-E(x)),\hspace{0.2cm} E\in \mathbb{C}
\end{eqnarray}
where x is any real variable, for example, position or fields (though in this paper we refer to it as position), one can look at the rate of change of the imaginary part of the action $Im(E)\equiv E_{im}(x)$ as
\begin{eqnarray}
\frac{\partial E_{im}(x)}{\partial x_j} \equiv F_j(x) \label{eq:force}
\end{eqnarray}
where we have defined a vector $F_j$ for the derivative of the imaginary part of the action. Though we call E an action, it can be any complex function, for instance, in the complex time direction in quantum mechanics, it will be the energy. In an N-dimensional space, there exist N-1 vectors that are orthogonal to the vector F. This means that at each point in space, there is only one direction in which the imaginary part is changing. The direction of this will depend on the position x.

 In theories with a sign problem, a way to fight the sign problem is to make every sample of $I$ as small as possible. The reason for this is that if you have a sum of +1 and -1, but the sum is highly oscillating between the 2 values, and you know the result should be $10^{-19}$, then you need a huge amount of samples to get an average that small. On the other hand, if you can change your sampling such that you only sample values of size $10^{-18}$, then one needs far fewer samples to get a precise result. One approach that uses this is steepest descent methods, which approximate paths around thimbles \cite{Cristoforetti:2012su,Alexandru:2016gsd}. These paths are chosen such that the imaginary part of the action stays constant, while the real part is always increasing, thus making all the oscillating samples much smaller. Another important observation is that many oscillating integrals can be well approximated by local solutions around fixed points, i.e. when $\frac{\partial E(x)}{\partial x_j} = 0$ for all $j$. 

The approach we propose here uses a different strategy to achieve a similar result, i.e. making values as small as possible and sampling mostly around fixed points, or at least where the imaginary part of the action is not changing significantly.

We propose to do a partial integration in a 1-dimensional subspace along paths given by

\begin{eqnarray}
\frac{dx_j}{d\tau} &=& F_j(x) = \frac{\partial E_{im}(x)}{\partial x_j}  \label{eq:path}
\end{eqnarray}
where $\tau$ is a fictitious time. Partial integration has previously been used in \cite{Kieu:1993ey,Bloch:2011jx} to tackle the sign problem.

\begin{figure}[h]
\begin{center}
  \includegraphics[width=8cm]{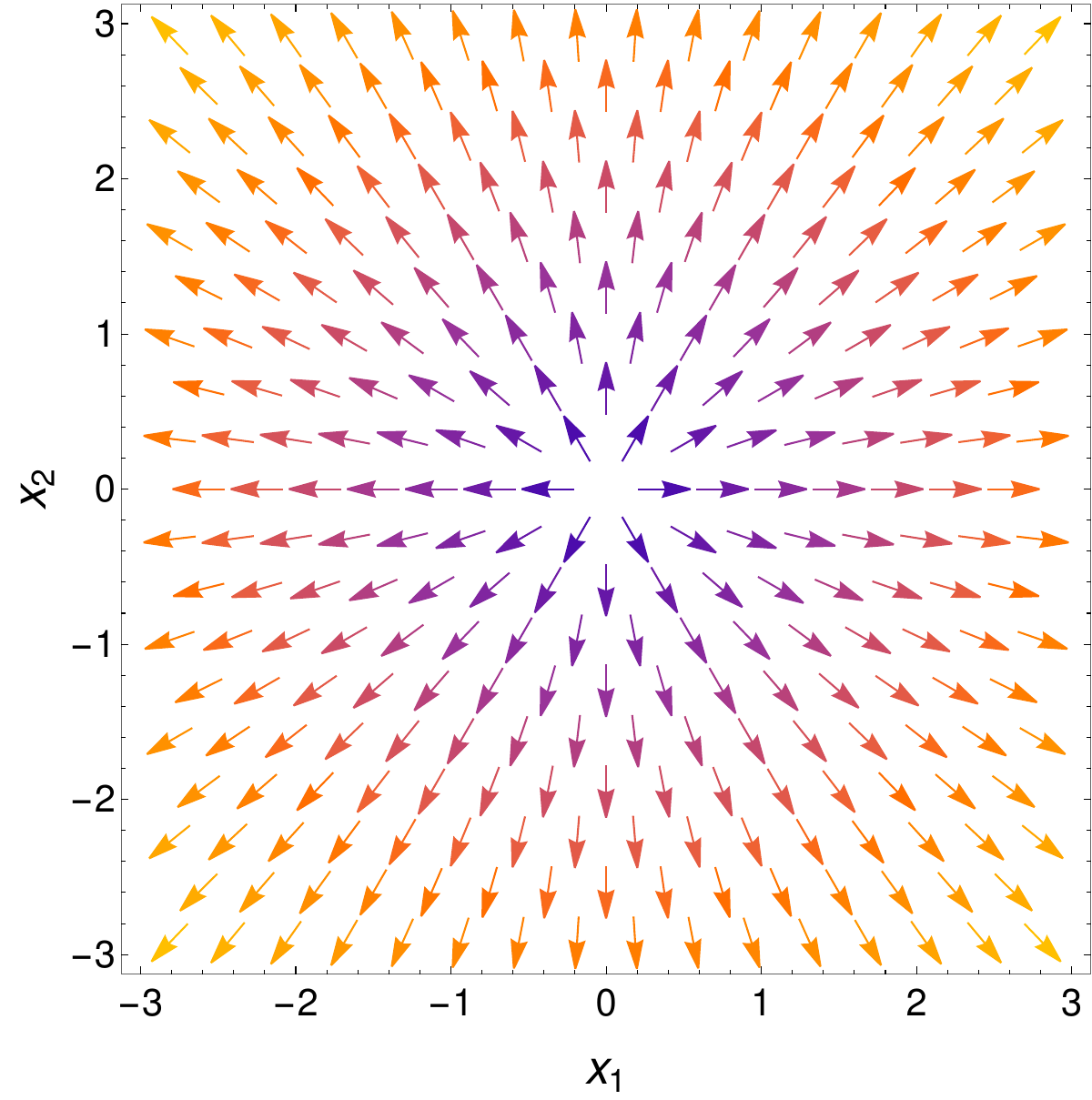}
   \caption{Vector field showing the direction of the paths $\frac{\partial E_{im}(x)}{\partial x_j} \equiv F_j(x) = 2 x_j$ for the example action $E=i(x_1 ^2 +x_2 ^2)$ as defined in eq. (\ref{eq:force}) and (\ref{eq:path}). }
   \label{fig:vector}
   \end{center}
\end{figure}

For example, if $E=i(x_1 ^2 +x_2 ^2)$, we get $F_j=2x_j$. From figure \ref{fig:vector} we see that all vectors point away from the fixed point at $x_1=x_2=0$, and we see that the distance between 2 paths will not stay the same, since the vectors are not parallel.  We need to keep these 2 observations in mind. From the perspective of eq. (\ref{eq:path}), a fixed point is when $\frac{\partial E_{im}(x)}{\partial x_j} = 0$, though when the action is completely imaginary, this definition and the derivative of the full action are equivalent.

\begin{figure}[h]
\begin{center}
  \includegraphics[width=9cm]{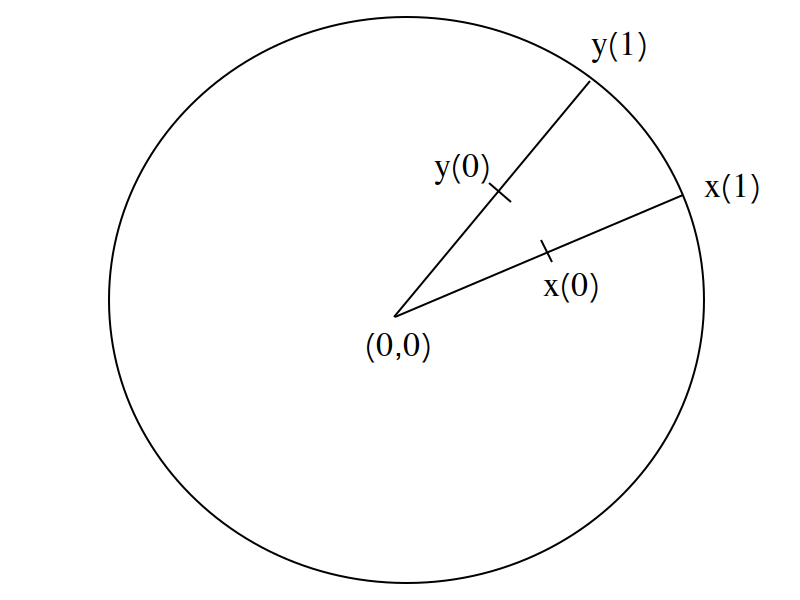}
   \caption{Sketch of different lines, based on the 2d example $F_j=2x_j$, obtained from initial conditions $x(\tau=0)$ and $y(\tau=0)$. We see how the distance between the 2 points increases for larger $\tau$. }
   \label{fig:vol}
   \end{center}
\end{figure}

The full path explored from eq. (\ref{eq:path}) in both positive and negative $\tau$ is here referred to as a line $I_O(x)$, where O refers to an observable measured along the path and x is the starting position from which the line was found. We wish to rewrite the original integral $I$, such that an integral over these lines is equal to the original integral
\begin{eqnarray}
I(O) &=& \int d^N x I_O(x) \label{eq:intFull}
\end{eqnarray}
with the expectation value of the observable O calculated as
\begin{eqnarray}
\langle O \rangle &=& \frac{I(O)}{I(1)} =  \frac{\int d^N x I_O(x)}{\int d^N x I_1(x)} \label{eq:intO}
\end{eqnarray}
 Along these lines, the imaginary part of the action $E$ will be changing rapidly. The central point of the proposed strategy is to carry out the integration over this highly oscillating part of the integral explicitly, for instance with a high order ordinary differential equation solver. This does require a high level of precision, which increases as the sign problem gets worse. For instance, in the examples shown later in figure \ref{nt12_sigma1_O2}, the precision needed was in the most oscillating lines as low as $10^{-20}$, though typically around $10^{-11}$ when closer to fixed points, where there are fewer oscillations.  The integral over the entire line, obtained from a starting point $x_0=x(0)$ following eq. (\ref{eq:path}), will be given by
\begin{eqnarray}
I_O(x_0) &=& \int _{-\infty} ^\infty  O(x(s)) \exp\left[-E(x(s))\right]V_{rel}(s) ds \label{eq:int} \\
     &=& \int _{-\infty} ^\infty O(x(\tau))  \exp \left[-E(x(\tau))+\sum _j \int _0^\tau \frac{\partial ^2 E_{im}(x(\tau'))}{\partial ^2 x_j}d\tau ' \right]|F(x_0)| d\tau   \nonumber
\end{eqnarray}
where $x(\tau)$ is obtained by following the defined path in both positive and negative $\tau$. s is the distance traveled along the line. Using eq. (\ref{eq:path}) the distance traveled s can be calculated as
\begin{eqnarray}
\frac{ds}{d\tau} &=& |F(x(\tau))| = \sqrt{\sum _j F_j(x(\tau)) ^2}
\end{eqnarray}
 We want to integrate over the entire line defined by the path from eq. (\ref{eq:path}), s therefore goes from -infinity to +infinity. $V_{rel}(s)$ is the relative volume factor as a function of s normalized to 1 at $s=0$, due to the path diverging or converging with other paths. To understand this effect, let's think back to the example of $F_j=2x_j$. As shown in figure \ref{fig:vol}, a point starting at $x(0)$ will at $\tau =1$ have moved out to $x(1)$. Another point $y$ starts at the same distance to the origin $(0,0)$, but with a different angle. At $y(1)$ the angle to $x(1)$ is the same as at $\tau=0$, but the distance has increased. It is the change of this distance that we call $V_{rel}$ and will be given as a $N-1$ dimensional volume factor. For the simple example, we find that $V_{rel}=\frac{|x(\tau)|}{|x(0)|}$, which is the factor obtained by going to radial coordinates.  

  The factor $V_{rel}$ has to be included to make sure that each point is counted the correct amount of times, since if we calculate the line integral not from $x_0=x(0)$ but from $x_0=x(1)$ there is a $V_{rel}=\frac{|x(1)|}{|x(0)|}$ higher chance to be sitting at $x(1)$ compared to $x(0)$. Including the volume factor $V_{rel}$ takes care of this difference. In the second line of eq. (\ref{eq:int}) we have rewritten $V_{rel}(s(\tau))$ as an integral using the formula in eq. (\ref{eq:det}) as shown below. The difference between the 2 paths arises from the difference in the vector F at different locations. The volume element is thus changed by 
\begin{eqnarray} 
\left[ dx_i(x+\epsilon _2 v_j)-dx_i(x)\right]/\epsilon_2=\epsilon\left[F_i(x+\epsilon _2 v_j)-F_i(x)\right]/\epsilon _2  &=& \epsilon \frac{\partial ^2 E_{im}(x)}{\partial x_i \partial x_j} v_j +O(\epsilon_2) \nonumber \\
\frac{V_{rel}(\tau + \epsilon)}{V_{rel}(\tau)} = \det\left[I_{ij}+\epsilon \frac{\partial ^2 E_{im}(x(\tau))}{\partial x_i \partial x_j}\right] &=& 1+\epsilon \sum _j \frac{\partial ^2 E_{im}(x(\tau))}{\partial ^2 x_j} \nonumber \\
   &&+ O(\epsilon ^2) \label{eq:deltaV}
\end{eqnarray}
where $v$ is chosen to be a unit vector in the j'th direction, $\epsilon$ and $\epsilon_2$ are infinitesimal changes in $\tau$ and position $x$ respectively and $I_{ij}$ is the identity matrix. Eq. (\ref{eq:deltaV}) can thus be solved as an ordinary differential equation with
\begin{eqnarray}
\frac{d \log(V_{rel}(\tau))}{d\tau} &=& \sum _j \frac{\partial ^2 E_{im}(x(\tau))}{\partial ^2 x_j} \label{eq:det}
\end{eqnarray}
along side solving for the path $x(\tau)$. 

The factor $|F(x_0)|$ in eq. (\ref{eq:int}) arises from the change of coordinates from s to $\tau$, while the change of this factor as a function of $\tau$ is included in the change of $V_{rel}$ in eq. (\ref{eq:det}), since this equation also includes the change to the infinitesimal unit vector along the path of integration.

Let's see how the line integral works for the example of $E=i(x_1^2+x_2 ^2)$. We have $\frac{d x_j}{d \tau} = 2 x_j$ and $\sum _j \frac{\partial ^2 E_{im}(x(\tau))}{\partial ^2 x_j} = 4$. Solving for $x_j$ and using eq. (\ref{eq:int}) gives
\begin{eqnarray}
x_j &=& x_{0j}e^{2\tau}  \\
I_O(x_0) &=& \int _{-\infty}^\infty O(\tau) \exp( -i(x_{01}^2+x_{02}^2) e^{4\tau}+4\tau) 2\sqrt{x_{01}^2+x_{02}^2} d\tau
\end{eqnarray}
where $x_{0j}=x_j(\tau=0)$. We can then redefine $x_{01}^2+x_{02}^2 = r_0 ^2$ and use that the radius is given by $r^2=x_1^2+x_2^2=(x_{01}^2+x_{02}^2) e^{4\tau}$, or $4\tau = 2log(r/r_0)$ and $\frac{d\tau}{d r} = \frac{1}{2r}$, which then gives
\begin{eqnarray}
I_O(x_0) &=& \int _{-\infty}^\infty O(\tau) \exp( -ir^2+2\log(r/r_0)) 2r_0 d \tau   \\
I_O(r_0,\theta) &=& \int _{0}^\infty O(r,\theta) \exp( -ir^2+2\log(r/r_0)) (r_0/r) dr \\
&=& \int _{0}^\infty O(r,\theta) \exp( -ir^2) (r/r_0) dr
\end{eqnarray}
which is the radial integration of the original integral $I(O)$, weighted with the factor $1/r_0$. Using eq. (\ref{eq:intFull}) we can then recover the original integral by changing to radial coordinates
\begin{eqnarray}
I(O) &=& \int I_O(r_0,\theta) r_0 dr_0 d\theta \\
 &=& \int O(r,\theta) \exp( -ir^2) (r/r_0) r_0 dr  dr_0 d\theta \\
 &=& \int  O(r,\theta) \exp( -ir^2) r dr d\theta \int dr_0 
\end{eqnarray}
which gives the original integral times the extra factor of $\int dr_0 $, which is the length of the lines, which is how many times we have double counted the points. By starting from a finite length line and then taking the limit to infinity, one can see that all lines will have the same length, and the factor $\int dr_0 $ will therefore average out for any observable. Still, that does require one to make an infinitely long integral, which is difficult to do numerically. Instead, we show below a way to limit the range of integration, which also makes the factor a finite constant.

\begin{figure}[h]
\begin{center}
  \includegraphics[width=11cm]{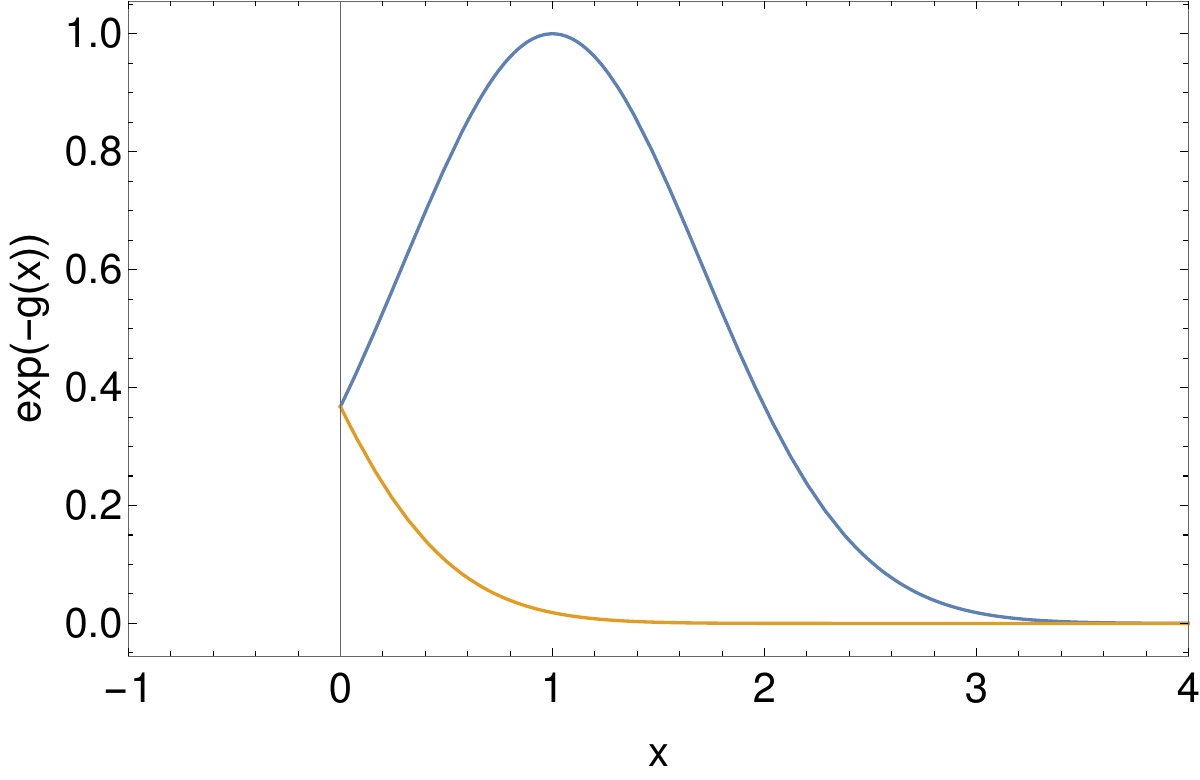}
   \caption{Example of cutoff function $\exp(-g(s))=\exp(-s^2)$ along coordinate x for a line starting at $x=1$ with a fixed point at $x=0$. The blue line is before hitting the fixed point and the yellow line is after. At $x=1$, 2 values of distance $s$ exist, $s=0$ (blue) and $s=-2$ (yellow). }
   \label{fig:hit}
   \end{center}
\end{figure}

In the definition of the line integral eq. (\ref{eq:int}), one has to integrate from $s=-\infty$ to $s=+\infty$. This is way too expensive to do numerically. We also expect, and have seen from simple examples, that the main contribution sits close to the fixed points. We therefore want to limit the range of the line integral in eq. (\ref{eq:int}), while still fulfilling eq. (\ref{eq:intO}), i.e., still being exact. We do this by introducing an extra integral along the line over a variable we will call y

\begin{eqnarray}
1 &=& constant \cdot \int _{-\infty} ^\infty \exp(-g(y)) dy  \\
\int _{-\infty} ^\infty f(x) dx &=&  constant \cdot \int _{-\infty} ^\infty f(x)\exp(-g(y)) dx dy \\
&=&  constant \cdot \int _{-\infty} ^\infty f(x+y)\exp(-g(y)) dx dy \nonumber
\end{eqnarray}
where we have shifted the $x$ integral to $x+y$. $f(x)$ will be the entire integrand of eq.  (\ref{eq:int}) and g is an arbitrary cutoff function, with the restriction that the integral has to converge. We want to perform the line integral of y and use x (alongside all other possible paths) as the initial position for the line integral. The integral over x can then be performed afterward, for instance, using Monte-Carlo methods on $x$. This does require the integral to go from -infinity to +infinity, but this can be solved in case we hit a fixed point $F_i(x) =0$ in the following way
\begin{eqnarray}
\int _0 ^\infty f(s) ds &=& \frac{1}{2}(\int _0 ^\infty f(s) ds +\int _0 ^\infty f(s) ds)  \\
                        &=& \frac{1}{2}(\int _0 ^\infty f(s) ds -\int _0 ^{-\infty} f(-u) du) \\
                        &=& \frac{1}{2}\int _{-\infty} ^\infty H(s) ds
\end{eqnarray}
where we defined $H(s)$, such that $H(s)=f(s)$, if s is positive, and $H(s)=f(-s)$ if s is negative. We do get a factor of a half, but as long as our cutoff function g is symmetric, we do not have to worry about the 2 contributions, since starting at s and -s will give the same result. This means that if one hits a fixed point, one simply reverses direction and keeps integrating. We show an example of the suppression factor coming from $\exp(-g(s))=\exp(-s^2)$ in figure \ref{fig:hit}. The distance traveled along the path s from an initial point on the line $x_0$, will keep increasing (or decreasing) along the entire path, even after having hit the fixed point. 

We thus end up with a line integral of the form
\begin{eqnarray}
I_O(x_0) &=& \int _{-\infty} ^\infty O(x(s))  \exp\left[-E(x(s))-g(s)\right]V_{rel}(s) ds \\
     &=& \int _{-\infty} ^\infty O(x(\tau))  \exp \left[-E(x(\tau))-g(s(\tau))+ \sum _j \int _0^\tau \frac{\partial ^2 E_{im}(x(\tau'))}{\partial ^2 x_j}d\tau' \right]|F(x_0)| d\tau \nonumber
\end{eqnarray}
g(s) should be chosen such that it cuts off the integral from going too far out, but should not cut off the integral too quickly, in order to allow the oscillatory behavior of the integrand to cancel out.
From the observation
\begin{eqnarray}
\sqrt{\frac{a}{\pi}} \int_{-\infty} ^\infty dx e^{-ax^2+bx} &=&  e^{b^2/(4a)} \label{eq:suppress}
\end{eqnarray}  
using $g(s) = as^2 = s^2/\sigma ^2$ can give a large suppression factor when b is mostly imaginary, both positive and negative. $\sigma$ will have to be chosen such that it suppresses the imaginary part enough, without wasting computer time. A larger $\sigma$ means that the highly oscillating regions will be more suppressed.

To conclude, in the strategy laid out above, we have transformed the original integral into an integral over line integrals instead. The line integrals are expressed in a form that can be solved using ordinary differential equation methods. A residual sign problem will exist, but it will be highly reduced.

\section{Implementation}
In order to implement the line integral, we write the problem as an ordinary differential equation such that 
\begin{eqnarray}
F_j(x) &=& \frac{\partial  E_{im} }{\partial x _j}\\
\frac{dx_j}{d \tau} &=& F_j(x)\\
\frac{ds}{d \tau} &=& \sqrt{\sum_ j  F_j(x) ^2} \\
\frac{dJ}{d \tau} &=& \sum _j \frac{\partial ^2 E_{im} }{\partial ^2 x _j} \\
\frac{dI_O}{d \tau} &=&  O(x(\tau)) e^{-E(x(\tau))-g(s)+J}|F(x_0)|
\end{eqnarray}
where we have defined $J=\log(V_{rel}(\tau))$. $|F(x_0)|$ can be absorbed into the initial conditions of $J$ as $J(0)=\log(|F(x_0)|)$. $g(s)$ can in principle be any symmetric function, but in this paper we will use $g(s) = (s/\sigma)^2$.

 In this form, we can use the existing libraries for solving ordinary differential equations. We have implemented our strategy using the Julia language using the DifferentialEquations.jl package \cite{rackauckas2017differentialequations}. Since we are dealing with highly oscillating integrals, we have found that a high order interpolation is required, and that the best performance is achieved with the method of DP8 (Hairer's 8/5/3 adaptation of the Dormand-Prince Runge-Kutta method (7th order interpolant)). The high precision integration takes care of cancellations, it is therefore important to make sure that the relative precision does not become smaller than the digits stored in the chosen precision, while at the same time increasing precision when necessary. Initial conditions are for $\tau=0$ and we need to evaluate $I_O(\infty)-I_O(-\infty)$.

If a fixed point is hit ($F_j=0$ for all j), the direction of integration should be reversed, such that the path will move back up along the path it came. This will not give the same contribution, as s will keep increasing (decreasing) along the path. An easy way to implement this is to reverse the sign of $d\tau$, $s$ and $I_O$ when the fixed point is reached, and then keep integrating. Just remember to flip the sign of $I_O$ again at the end.

We have included a Julia file together with this publication, with an implementation for the problem discussed in the next section \cite{code}.

\section{Results on $x^4$ Schwinger-Keldysh contour}
\begin{figure}[h]
\begin{center}
  \includegraphics[width=8cm]{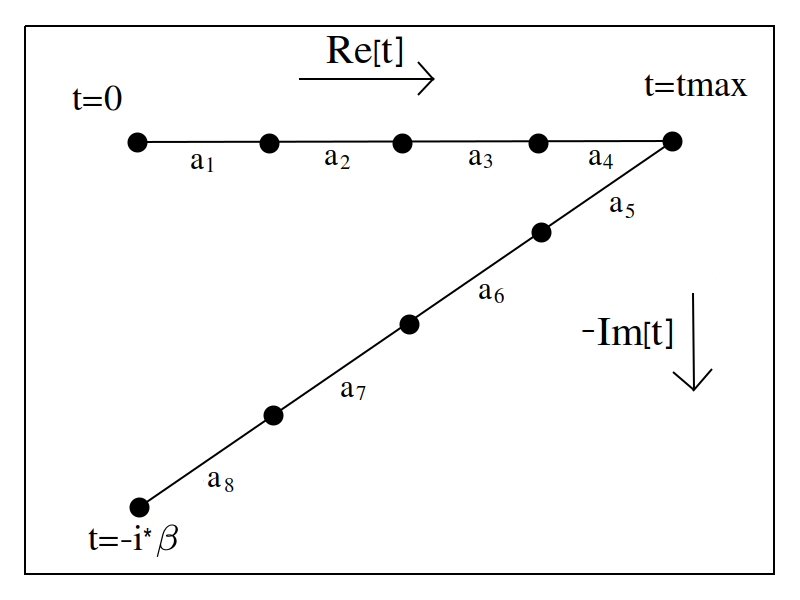}
   \caption{Contour for solving the finite temperature path integral. The contour is shown with 8 points for simplicity (the first and last point are the same point), though the examples use between 16 and 28. }
   \label{contour}
   \end{center}
\end{figure}

To compare with previous results of other techniques, such as Complex Langevin \cite{Berges:2006xc,Alvestad:2021hsi} and Lefschetz-thimbles \cite{Alexandru:2016gsd}, we work with a 1d quantum mechanical anharmonic  oscillator with a $x^4$ potential in real time, finite temperature. The time evolution we try to solve is
\begin{eqnarray}
\langle O \rangle &=& Tr(e^{-\beta H}xe^{-it H} xe^{it H})/Tr(e^{-\beta H})\\
H&=& \frac{p^2}{2}+\frac{x^2}{2}+\frac{\lambda x^4}{4!}
\end{eqnarray}
which can be solved by discretizing x and treating everything as finite size matrices. We will compare the solution obtained from discretizing the Hamiltonian, with the results obtained from the line integral method, which will be applied to the path-integral expressed as
\begin{eqnarray}
\langle O \rangle &=& \int d^Nx\exp \left(\sum _{j=1} ^Ni[\frac{(x_j-x_{j+1})^2}{2a_j} -\frac{(a_j+a_{j-1})}{2}(\frac{x_j^2}{2}+\frac{\lambda x_j^4}{4!})] \right) O(x)/Z  \nonumber \\
E &=& -\sum _{j=1} ^Ni[\frac{(x_j-x_{j+1})^2}{2a_j} -\frac{(a_j+a_{j-1})}{2}(\frac{x_j ^2}{2}+\frac{\lambda x_j^4}{4!})] \\
\frac{\partial E_{im}}{\partial x _j} = F_j(x) &=& -Im\left( i[\frac{(x_j-x_{j+1})}{a_j}+\frac{(x_j-x_{j-1})}{a_{j-1}} -\frac{(a_j+a_{j-1})}{2}(x_j+ \frac{\lambda x_j^3}{3!})]  \right) \\
\frac{\partial ^2 E_{im}}{\partial ^2 x _j} &=& -Im\left( i[\frac{1}{a_j}+\frac{1}{a_{j-1}} -\frac{(a_j+a_{j-1})}{2}(1+\frac{\lambda x_j^2}{2})]  \right)
\end{eqnarray} 
where $a_j$ parameterizes the Schwinger-Keldysh contour, which in our case is chosen as shown in figure \ref{contour} and Z is the partition function (the path-integral without the observable). The lattice is periodic, such that $x_j=x_{j+N}$. The observable of interest $O$ in the path-integral is given as $O = x(0)x(t)$.  

We indicate the number of points on the forward path as $N_+$ and for the tilted backward path as $N_-$. For example, for $N_+=N_-=10$ and $\beta = t_{max}=1.0$ we have that for $j$ in 1 to 10, $a_j=0.1$ while for $j$ in 11 to 20, $a_j = -0.1-0.1i$.

 We solve the path integral by calculating the line integrals as explained in the previous sections to obtain $I_1(x)$ and $I_O(x)$. During the simulation, we increase precision such that the result is at least $10^6$ times larger than the tolerance. We will then sample on $|I_1(x)|$ using the Metropolis–Hastings algorithm and compute the expectation value of the observables $O$ as 

\begin{eqnarray}
\langle O \rangle &=& \frac{\int d^N x I_O(x)}{\int d^N x I_1(x)} = \frac{\int d^N x |I_1(x)|\times I_O(x)/|I_1(x)|}{\int d^N x |I_1(x)|\times I_1(x)/|I_1(x)|} \\
 &=& \frac{\sum _j I_O(x_j)/|I_1(x_j)|}{\sum _j I_1(x_j)/|I_1(x_j)|}
\end{eqnarray}

where the subscript $O$ indicates the observable included in the line integral and j indicates the j'th measurement. We also define $\langle 1 \rangle$ as the measurement of the average phase (i.e. without dividing by itself)
\begin{eqnarray}
\langle 1\rangle&=&\frac{\sum _j I_1(x_j)/|I_1(x_j)|}{\sum _j 1}
\end{eqnarray}

\begin{figure}[h]
\begin{center}
  \includegraphics[width=5cm]{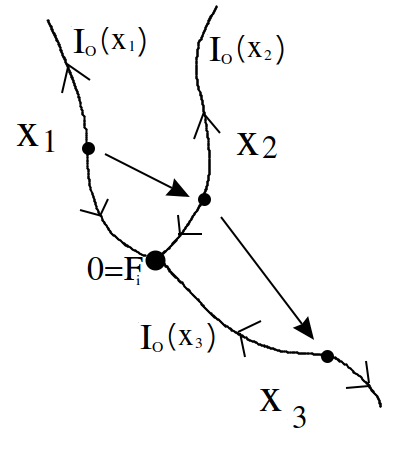}
   \caption{Sketch of the sampling process for lines. A line $I_1(x_1)$ is found starting from position $x_1$. A new position $x_2$ and line $I_1(x_2)$ are then accepted or rejected based on  $\left|\frac{I_1(x_2)}{I_1(x_1)}\right| $. This process repeats as many times as needed for good enough statistics. The arrows show how the initial position changes during the Monte-Carlo process. The lines show the paths found from each initial position. A fixed point $F_i(x) =0$ is also indicated. }
   \label{sampling}
   \end{center}
\end{figure}

The process for sampling the observables using the Metropolis–Hastings algorithm is thus the following. We start at an initial position $x_1$ where we calculate the line integrals $I_1(x_1)$ and $I_O(x_1)$. We then randomly jump to a new position $x_2$, at which $I_1(x_2)$ and $I_O(x_2)$  are calculated and a random number $r$ between 0 and 1 is drawn from a random generator. We then accept the new position $x_2$ if $\left|\frac{I_1(x_2)}{I_1(x_1)}\right| > r$ and otherwise reject the new position $x_2$. This process then repeats. We sketch this process in figure \ref{sampling}.

\begin{figure}[h]
\begin{center}
  \includegraphics[width=12cm]{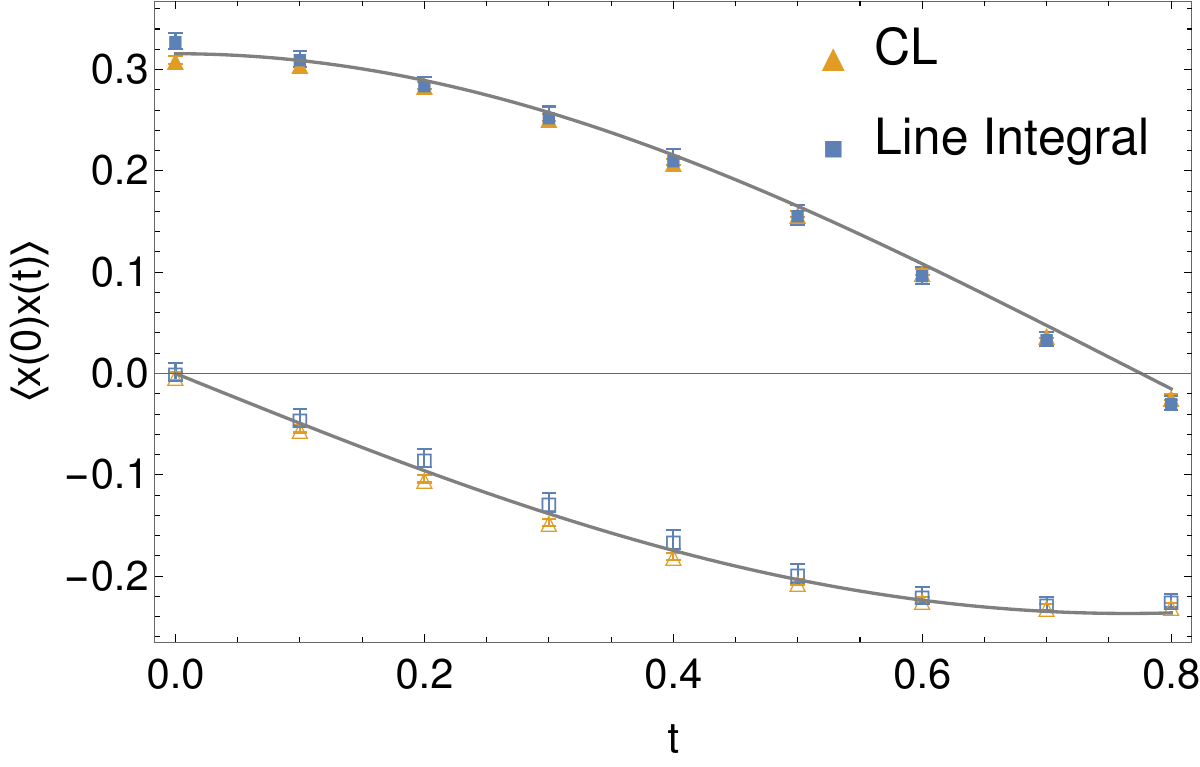}
   \caption{Real (Filled square) and imaginary (Open square) part of $\langle x(0)x(t) \rangle$ for $t_{max}=0.8$, $\beta=1$, $\lambda=4!$ with $N_+=8$, $N_-=12$ and $\sigma=1$. 100 streams of 90k points were used. Grey lines show the solution from discretizing the Hamiltonian. We compare to CL results (yellow triangles) shared from work done in \cite{Alvestad:2021hsi}. }
   \label{nt8_sigma1_O2}
   \end{center}
\end{figure}

\begin{figure}[h]
\begin{center}
  \includegraphics[width=12cm]{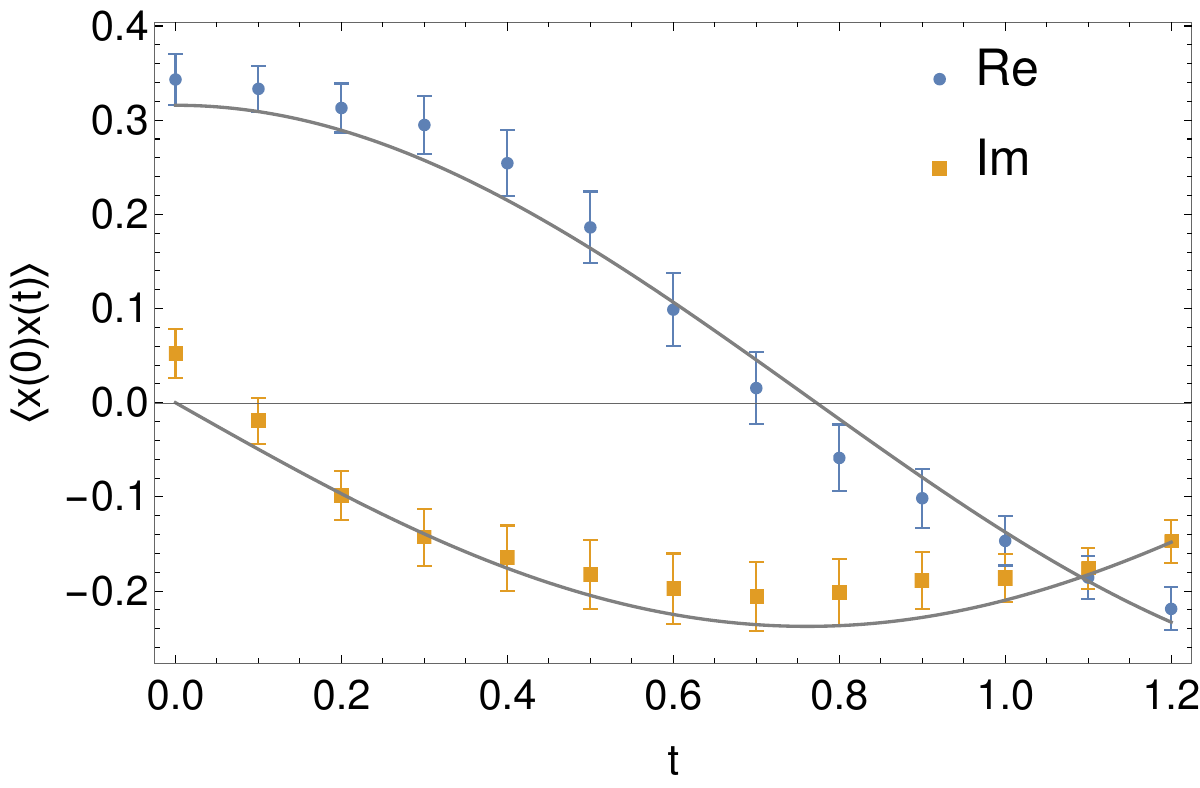}
   \caption{Real (blue circle) and imaginary (yellow square) part of $\langle x(0)x(t)\rangle$ for $t_{max}=1.2$, $\beta=1$, $\lambda=4!$  with $N_+=12$, $N_-=16$ and $\sigma=1$. 1000 streams of 100k points were used. Grey lines show the solution from discretizing the Hamiltonian.}
   \label{nt12_sigma1_O2}
   \end{center}
\end{figure}

\begin{figure}[h]
\begin{center}
  \includegraphics[width=12cm]{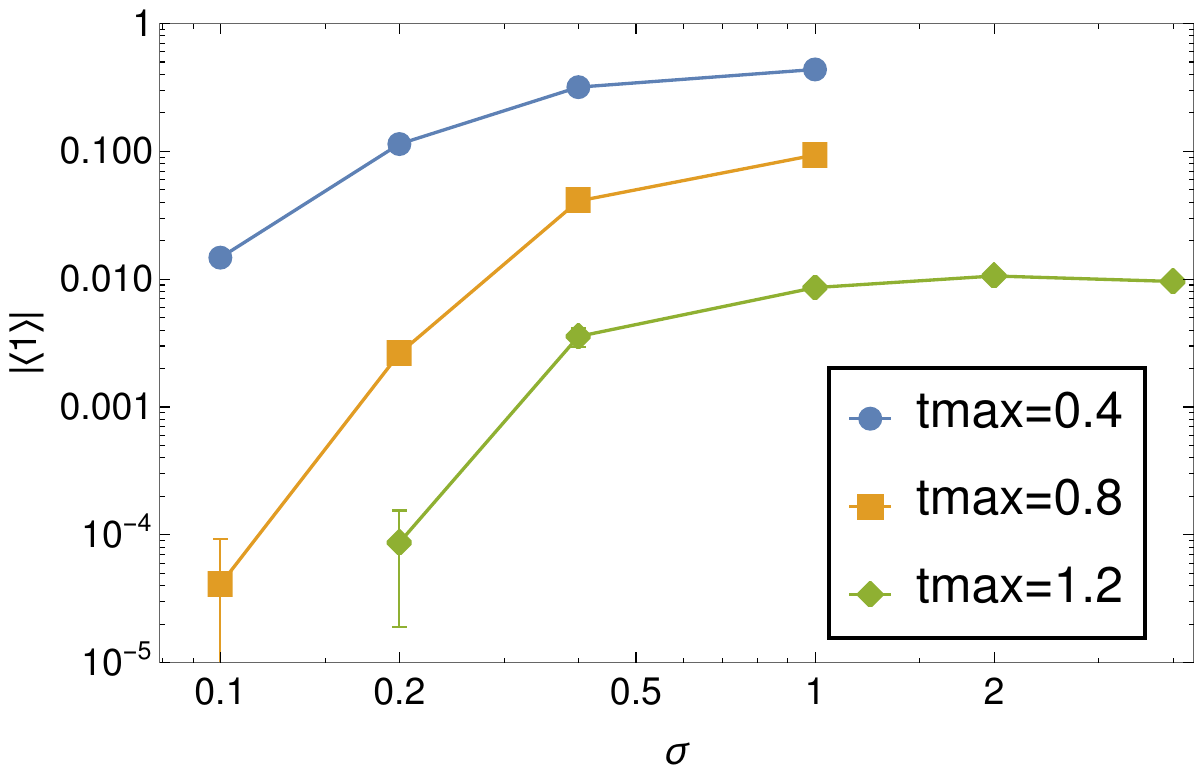}
   \caption{Absolute value of average sign $\langle 1\rangle$, for 3 different $t_{max}$ and different cutoffs $\sigma$ for cutoff function $\exp(-g(s))=\exp(-(s/\sigma)^2)$. The average sign is expected to go to exactly 0 for $\sigma = 0$. $\sigma=0$ corresponds to sampling with points.}
   \label{fig:sign}
   \end{center}
\end{figure}

Following previous works, we use $\beta = 1.0$ and $\lambda=4!$, which typically allow for measuring up to a real time of order 1. We show examples for $t_{max}=0.8, 1.2$ in figure \ref{nt8_sigma1_O2} and \ref{nt12_sigma1_O2} respectively.
Complex Langevin without any modification gets into problems around $t_{max}=0.8$ \cite{Berges:2006xc,Alvestad:2021hsi}, due to the problem with converging to the wrong solution, while  Lefschetz-thimbles are able to reach $t_{max}=2.0$ \cite{Alexandru:2016gsd}, though with much rougher spacing $a_j$. As shown in figure \ref{nt12_sigma1_O2}, we are currently able to reach a maximum time of $t_{max}=1.2$, at which point the error bars start to become too large, due to the absolute value of the average sign becoming very small.  We have in these calculations used the cutoff function $g(s)=(s/\sigma)^2$. We found that $\sigma \sim 1$ works well for this problem. While larger values of $\sigma$ do decrease the sign problem (increases the size of the average sign), the gain from this  was less than the loss due to increased computing time above $\sigma = 1$ (see figure \ref{fig:sign} and table \ref{table:data}). $\sigma< 1$ however quickly makes the sign problem stronger as shown in figure \ref{fig:sign} and table \ref{table:data}, and we expect $\sigma =0$ to have an average sign of $\langle 1 \rangle = 0$, due to the real time evolution having a completely complex action. $\sigma = 0$ corresponds to the standard sampling method of sampling with points, and the relative improvement over point sampling is therefore infinite for real time dynamics like the explored 1d quantum mechanical anharmonic oscillator. From figure \ref{fig:sign} we see that as $t_{max}$ increases, the residual sign problem increases, which increases the required statistics quickly. 
 
\begin{table}[h]
\begin{center}
\begin{tabular}{ |c|c|c|c|c|c|c| } 
 \hline
  $t_{max}$ & $N_+$  & $N_-$  &  $\sigma$ & $|\langle 1\rangle |$ & Measured & time/measurement \\ 
 \hline
 0.4 & 4 & 12 & 0.1 & $0.0147 \pm 0.0008$ & 9M & 0.0024s \\ 
 \hline
 0.4 & 4 & 12 & 0.2 & $0.114 \pm 0.001$ & 9M & 0.0028s \\ 
 \hline
 0.4 & 4 & 12 & 0.4 & $0.318 \pm 0.002$ & 9M & 0.0037s \\ 
 \hline
 0.4 & 4 & 12 & 1 & $0.44 \pm 0.02$ & 900k & 0.011s \\ 
 \hline
  0.8 & 8 & 12 & 0.1 & $4.1\times 10 ^{-5} \pm 5.1\times 10 ^{-5}$ & 900M &  0.0040s \\ 
   \hline
   0.8 & 8 & 12 & 0.2 & $2.6\times 10 ^{-3} \pm 2.4\times 10 ^{-4}$ & 90M &  0.0045s \\ 
   \hline
  0.8 & 8 & 12 & 0.4 & $0.041 \pm 0.001$ & 9M &  0.0068s \\ 
   \hline
  0.8 & 8 & 12 & 1 & $0.094 \pm 0.001$ & 9M &  0.017s \\ 
 \hline
  1.2 & 12 & 16 & 0.2 & $8.7\times 10 ^{-5} \pm 6.8\times 10 ^{-5}$ & 1000M & 0.015s\\ 
 \hline
  1.2 & 12 & 16 & 0.4 & $0.0036 \pm 0.0006$ & 100M & 0.025s\\ 
   \hline
  1.2 & 12 & 16 & 1 & $0.0086 \pm 0.0003$ & 100M & 0.06s\\ 
 \hline
 1.2 & 12 & 16 & 2 & $0.0106 \pm 0.001$ & 10M  & 0.13s\\ 
 \hline
 1.2 & 12 & 16 & 4 & $0.0096 \pm 0.001$ & 10M & 0.33s\\ 
 \hline
\end{tabular}
\caption{Summary of parameters used for the different runs and the obtained average sign. The runs were done on 3 different CPUs, and time/measurement does therefore have some variation in it. \label{table:data}}
\end{center}
\end{table}

\section{Summary}
We have defined a class of line integrals, where the integral over all lines adds up to the integral from which the line integrals were derived.  The path of each line integral is obtained from an initial position $x_0$, which is then flowed to plus and minus infinity using the direction of change in the imaginary part of the action. This allows for cancellations in the oscillating integral, making regions of quickly changing imaginary action less likely. These line integrals reduce the sign problem significantly and make it possible to use standard Monte-Carlo methods on the initial position of the line $x_0$, where it would normally not be possible due to the complex action.

The line integrals can effectively be solved using modern ordinary differential libraries. We implemented this using the DifferentialEquations.jl package \cite{rackauckas2017differentialequations} in the Julia language. A Julia file  (jl) is attached to this publication, with code used to produce the results \cite{code}.

 We applied these line integrals to a 1d quantum mechanical anharmonic  oscillator with a $x^4$ potential in real time, finite temperature. We found for $\beta=1.0$, $\lambda=4!$ that we can simulate up to a real time of $t_{max}=1.2$, after which the residual sign problem becomes too strong. While not quite as large a $t_{max}$ as Lefschetz-thimbles \cite{Alexandru:2016gsd} methods (though our results use smaller lattice spacing $a_j$) or some improved Complex Langevin methods \cite{Alvestad:2022abf}, the real time extent is not far off  compared to the other methods and exceeds the real time extent where standard Complex Langevin fails. 
 
This work was done as a first attempt, but the line integrals still have extra freedom to play with, which hopefully can further reduce the sign problem, as is seen necessary at long real time extents.  For instance, the cutoff function $g$ is arbitrary, though chosen with a good reason, since for completely imaginary actions it creates an exponential suppression as seen in eq. (\ref{eq:suppress}), though it could be that other functions work better. A different approach is to change the paths of the line integral. While moving along lines defined by the direction of change in the imaginary part of the action makes sense due to the high oscillations along these paths, other possible choices could be better, especially in theories with an action that has large contributions from both the real and imaginary part. One could for instance try to move along lines of changing imaginary action that are orthogonal to the direction of change to the real part of the action.  A third approach could be to redefine the scale of the local dimensions. The locally fastest path down a mountain depends on the scale of the different directions. It could be that for interacting theories, like $x^4$, a better scaling would be where all directions are important, instead of the current situation, where the j'th direction for which $x_j ^4$ is largest quickly dominates, since the line will tend to flow only in the j'th direction.

\section*{Acknowledgements}
The author gladly acknowledges support by the Research Council of Norway under the FRIPRO Young Research Talent grant 286883. The author would like to thank Daniel Alvestad for helping set up the code in Julia, for discussion on how to optimize the Julia code and for providing the Complex Langevin result for comparison. The author would also like to thank Alexander Rothkopf and Daniel Alvestad for their input to the draft of this paper.

%%%%%%%%%%%%%%%%%%%%%%%%%%%%%%%%%%%%%%%%%
%%                                     									       %%
%%  So called Backmatter part starts here with acknowledgements		       %%
%%  funding information, and the bibliography.						       %%
%%                                     									       %%
%%%%%%%%%%%%%%%%%%%%%%%%%%%%%%%%%%%%%%%%%

\FloatBarrier

\begin{backmatter}

%\section*{Competing interests}
%  The author declares that he has no competing interests.

%\section*{Author's contributions}
%    \begin{itemize}
%        \item F. Last: conceptualization, implementation, writing
%    \end{itemize}

%%%%%%%%%%%%%%%%%%%%%%%%%%%%%%%%%%%%%%%%%
%%                                     									       %%
%%  Bibliography part starts here								       %%
%%                                     									       %%
%%%%%%%%%%%%%%%%%%%%%%%%%%%%%%%%%%%%%%%%%

\bibliographystyle{stavanger-mathphys}

%%%%%%%%%%%%%%%%%%%%%%%%%%%%%%%%%%%%%%%%%
%%                                     									       %%
%%  Specify your BibTeX bibliography file here or manually insert references  %%
%%                                     									       %%
%%%%%%%%%%%%%%%%%%%%%%%%%%%%%%%%%%%%%%%%%

\end{backmatter}

%%%%%%%%%%%%%%%%%%%%%%%%%%%%%%%%%%%%%%%%%
%%                                     									       %%
%%  End of the document										       %%
%%                                     									       %%
%%%%%%%%%%%%%%%%%%%%%%%%%%%%%%%%%%%%%%%%%

\end{document}